# S2DS: Physics-Based Compact Model for Circuit Simulation of Two-Dimensional Semiconductor Devices Including Non-Idealities

Saurabh V. Suryavanshi[1] and Eric Pop[1,2,3,*]

[1]*Dept. of Electrical Engineering, Stanford University, Stanford, CA 94305, U.S.A.*

[2]*Dept. of Materials Science and Engineering, Stanford University, Stanford, CA 94305, U.S.A.*

[3]*Precourt Institute for Energy, Stanford University, Stanford, CA 94305, U.S.A.*

We present a physics-based compact model for two-dimensional (2D) field-effect transistors (FETs) based on monolayer semiconductors such as $MoS_2$. A semi-classical transport approach is appropriate for the 2D channel, enabling simplified analytical expressions for the drain current. In addition to intrinsic FET behavior, the model includes contact resistance, traps and impurities, quantum capacitance, fringing fields, high-field velocity saturation and self-heating, the latter being found to play a strong role. The model is calibrated with state-of-the-art experimental data for n- and p-type 2D-FETs, and it can be used to analyze device properties for sub-100 nm gate lengths. Using the experimental fit, we demonstrate feasibility of circuit simulations using properly scaled devices. The complete model is implemented in SPICE-compatible Verilog-A, and a downloadable version is freely available on the nanoHUB.org.

[*]Contact: epop@stanford.edu

I. INTRODUCTION

There has been growing interest in two-dimensional (2D) semiconductor devices and circuits after recent developments in the fabrication and growth of 2D materials beyond graphene.[1] Layered semiconductors include transition metal dichalcogenides (TMD) like $MoS_2$ and so-called "X-enes" like phosphorene. The thickness of one *monolayer* (1L) of such 2D materials is defined as the layer spacing from X-ray diffraction of the bulk material, and it is typically less than 1 nm (for $MoS_2$, $t_{2D} = 6.15$ Å).[2] While 1L of some materials (e.g. phosphorene) have been difficult to isolate and measure due to the lack of stability in ambient conditions, most sulfur-based TMDs like $MoS_2$ are stable and can be grown as large-area monolayers with promising electrical properties.[3,4] Sub-nanometer channel thickness is expected to enable good gate control and to minimize short channel effects in field-effect transistors (FETs) with sub-10 nm gate lengths.[5,6] It also appears that the mobility (and therefore carrier mean free paths) of 2D materials are better preserved than those of bulk materials (e.g. Si) when the channel thickness is scaled below ~3 nm.[7] In addition, several simulation studies have recently suggested that 2D FETs could benefit from the larger band gap of 1L TMDs, reducing the off-state leakage current.[8,9]

Although there remains much room for improvement in 2D devices, nascent efforts are already underway to integrate them into simple circuits with few transistors.[10–12] With recent reports of wafer-scale uniform growth and improved mobilities,[3,4] as well as smaller contact resistance,[7] larger-scale circuits will soon become feasible. This push towards applications of 2D FETs requires a platform to evaluate such devices at the circuit and systems level. With many 2D materials being tested, there is also a growing requirement to analyze a larger material parameter space. In this context, Jiménez introduced a model that captured essential physics of ideal 2D FETs, but did not include non-idealities and parasitics present in realistic devices.[13] Other 2D FET models accounted for device electrostatics based on Poisson's equation with varying degrees of complexity, however without inclusion of high-field velocity saturation,[14] and fringing field or self-heating and temperature-dependent effects.[14,15]

In this paper, we describe the Stanford 2D Semiconductor (S2DS) transistor model, a physics-based and data-calibrated model to simulate circuits and systems made from 2D materials. S2DS has been implemented in Verilog-A (Ref. 16) and is freely available online.[17] The model description provided in this paper focuses on *monolayer* 2D materials because they are most promising from a scaling point of view, but S2DS can also treat few-layer FETs if properly calibrated. The paper is organized as follows. First, we derive an analytic expression for the drain current and calibrate the model with existing experimental data. Then, we describe additional effects, including high-field velocity saturation, self-heating, contact resistance, and fringing fields that are essential to understand the realistic behavior of 2D FETs. Importantly, self-heating is found to play a strong role, particularly as the quality of 2D materials improves, and the current densities achieved in experiments increase. The ultimate purpose of S2DS is to help the 2D community understand transistor measurements and to assist in designing further experiments. The SPICE-compatible model also enables design and benchmarking of large-scale circuits and systems composed of 2D devices with varying substrates of differing thermal properties (e.g. silicon versus glass or plastics).

## II. COMPACT MODEL

### A. Intrinsic I-V Characteristics

Figure 1 shows the schematic of a FET based on a monolayer 2D semiconductor, with its main parasitic components. For the sake of concreteness, a typical device structure considered in this work includes the 1L material on an insulator (e.g. the bottom oxide shown in Fig. 1), a top-gate oxide and gate metal stack on top. A finite underlap distance ($L_U$) appears between the edge of the top-gate and the source and drain contacts, respectively. The contacts themselves have a finite length $L_C$, which can play a role when this becomes comparable to the current transfer length.[7] Beneath the bottom oxide, the substrate can be conductive as a doped Si wafer (which is often used as a back-gate in experiments), or insulating like glass, quartz, or a flexible polymer.

To simplify the analysis, we present equations for n-type transistors (n-FET), considering electron transport in the channel and positive voltages ($V_{GS}$, $V_{DS} > 0$). Of course, these can be easily modified to simulate p-type transistors (p-FET), as is done for inverter simulations in the latter part of this study. We build on the model developed by Jiménez to derive the current-voltage ($I$-$V$) characteristics, with several extensions described below, particularly with respect to "extrinsic" transistor aspects, including self-heating, velocity saturation, and fringing fields.[13,18]

First, we include the effect of the band structure for charge calculation. For example, in most TMDs (including $MoS_2$) two conduction band valleys may participate in transport, one at the K point and the other halfway along the K and Γ points of the Brillouin zone, sometimes labeled the Q valley.[19–21] Figure 2a shows the schematic of this band structure, where $\Delta E_{KQ}$ is the energy separation between the K and Q conduction band valleys. We calculate the charge density as

$$n_{2D} = \int_0^\infty DOS_{2D}(E) \cdot f(E) dE\,, \tag{1}$$

where $f(E)$ is the Fermi-Dirac distribution and $DOS_{2D}(E)$ is the 2D density of states corresponding to the lowest band. The Fermi energy $E_F = qV_C$, where $V_C$ is the voltage across the quantum capacitance for the 2D channel and $q$ is the elementary charge. Simplifying the charge density expression, we obtain $n_{2D} = N_{2D} \ln(1 + \alpha)$ where $\alpha = \exp[(qV_C - E_0)/(k_B T)]$, $E_0 = E_G/2$, and

$$N_{2D} = \frac{k_B T g_K m_{effK}}{\pi \hbar^2} + \frac{k_B T g_Q m_{effQ}}{\pi \hbar^2} \exp\left(-\frac{\Delta E_{KQ}}{k_B T}\right) \tag{2}$$

where, $k_B$ is the Boltzmann constant, and $T$ is the average device temperature. We use the semiconductor mid-gap as our energy reference ($E = 0$) such that the conduction band energy is $E_0$ and the valence band energy is $-E_0$. $g_K$ and $g_Q$ are the degeneracy of the K and Q conduction valleys, respectively, and $m_{effK}$ and $m_{effQ}$ are their respective DOS effective masses. For MoS$_2$ $g_K = 2$, $g_Q = 6$,[14] $m_{effK} = 0.48 m_0$, and $m_{effQ} = 0.57 m_0$.[22] $\Delta E_{KQ}$ is the energy separation between K and Q conduction valleys (~0.11 eV for monolayer MoS$_2$).[19–21] For the sake of simplicity, we take the band gap ($E_G$) to be the same as the photoluminescence (PL) gap, which is ~1.85 eV for MoS$_2$ and ~1.65 eV for WSe$_2$. However, we note that the true electronic band gap can be affected by the dielectric screening environment, by strain, and proximity to grain boundaries.[20,23]

Along with the free charge carriers, impurities ($N_{Dop}$) and traps ($N_{it}$) also contribute to the total channel charge ($Q_{ch}$) as:

$$Q_{ch} = -q[N_{Dop} + N_{it} + n_{2D}]\,. \tag{3}$$

As shown in Fig. 2a, we model the interface traps as acceptors, situated at an effective energy $E_{it}$ below the conduction band, with an effective trap density $D_{it}$. To simplify the model, $D_{it}$ is assumed here to be a delta function in energy, but this approach could be generalized. At a particular bias, the number of trapped carriers ($N_{it}$) is given by,

$$N_{it} = \int_{-E_0}^{E_0} D_{it} f(E) dE = \frac{D_{it}}{1 + \exp(\frac{E_0 - E_{it} - qV_C}{k_B T})}\,. \tag{4}$$

The device electrostatics are guided by the distributed capacitive circuit model shown in Fig. 2(b).[13,24] The top and the back oxide capacitance are $C_t$ (= $\varepsilon_{OX}/t_{OX}$) and $C_b$ (= $\varepsilon_{BOX}/t_{BOX}$) respectively. $\varepsilon_{OX}$ and $\varepsilon_{BOX}$ are the dielectric constants, and $t_{OX}$ and $t_{BOX}$ are the oxide thicknesses of the top and bottom oxide, respectively. $C_q$ is the quantum capacitance and $C_{it}$ is the capacitance due to traps at the oxide-semiconductor interface, taken as a combination from both interfaces of the ultra-thin 2D channel. The quantum capacitance $C_q$ and the trap capacitance $C_{it}$ are given by

$$C_q = q\frac{dn_{2D}}{dV_C} = \frac{q^2 N_{2D}\alpha}{(1+\alpha)k_B T} \qquad (5a)$$

$$C_{it} = -q\frac{dN_{it}}{dV_C} = \frac{D_{it}q^2\alpha\beta}{k_B T(\alpha+\beta)} \qquad (5b)$$

where $\beta$ = exp[$-E_{it}/(k_B T)$].

$V_{GS}$-$V_{GS0}$ and $V_{BS}$-$V_{BS0}$ are internal voltages from the top- and the back-gate respectively. $V_{GS0}$ and $V_{BS0}$ are flatband voltages of the top- and back-gate, treated as fit parameters to the experimental data. (For example, if the top-gate metal workfunction is increased, $V_{GS0}$ will be higher, etc.) The total charge in the 2D channel ($Q_{ch}$, eq. 3), and the quantum potentials at the source ($V_{Cs}$) and the drain ($V_{Cd}$) are calculated iteratively as discussed in the Appendix A, including doping and trapped charges. We neglect the channel depletion capacitance because the channel thickness is less than 1 nm. We note that the effect of the fringing field from the drain through the BOX can also be incorporated in our model by including an additional capacitance between the drain node and the channel in the circuit shown in Fig. 2(b).[25] In thicker multi-layer channels, the depletion capacitance should be accounted for, in a similar manner as it is done for silicon-on-insulator (SOI) transistors.[26]

We solve for semi-classical drift transport to obtain an expression for the drain current ($I_D$ = -$I_S$) for all transistor operating regions. The semi-classical approach is appropriate even for 2D FETs near 10 nm channel length, as the present-day experimental mean free path in monolayer 2D semiconductors like $MoS_2$ is ~2 to 3 nm (see Fig. S10 in Supplement of Ref. 3). Similarly, our approach should hold for channel widths greater than 10-20 nm, for which edge scattering effects can be safely ignored. (And all experimental data for 2D semiconductors is typically taken on micron-width devices, to obtain larger current drives.) Thus:

$$I_D = \frac{\mu W}{L_G}\left[ N_{2D}k_BT\alpha + q^2\ln^2(1+\alpha) - \frac{N_{2D}q^2D_{it}\beta}{(C_t+C_b)(\beta-1)}\left( \frac{(1+\alpha)\ln(1+\alpha)}{\alpha+\beta} - \ln(\alpha+\beta) \right) \right]_{V_{Cd}}^{V_{Cs}}. \quad (6)$$

Here $\mu$ is the carrier mobility, $C_t$ and $C_b$ are the top and the bottom oxide capacitances per unit area, and other variables are defined earlier. We recall that $\alpha$ is a function of $V_C$, and thus $I_D$ is calculated as the difference of eq. 6 evaluated at $V_{Cs}$ and $V_{Cd}$. (The complete derivation is given in Appendix A.) Gate and diffusion currents are not included in the present version of the S2DS model, thus leakage power will be underestimated. Nevertheless, this could be a reduced component in TMD FETs, which have larger band gaps than semiconductors like Si and Ge.

When fitting to some (but not all) experimental data, we need to introduce a finite output resistance modeled by a fit parameter $\lambda$ as $I_{D,\text{eff}} = I_D(1 + \lambda V_{DS})$. However, sometimes $\lambda$ is not needed, especially when fitting the model against long channel back-gated MoS$_2$ FETs.[3] For fitting the model with experimental data on top-gate transistors, we used a finite value $0 < \lambda < 0.1$.[11,27] We note that the current saturation region is also influenced by device self-heating, which is taken into account self-consistently, as we will discuss below.

With these considerations, Fig. 3 displays the $I_D$ vs. $V_{GS}$ curve for a few trap densities and trap energies. For large trap densities, a part of the gate voltage is used to charge the traps. As a result, less voltage is available to induce mobile charges in the channel, leading to smaller drain currents [Fig. 3(a)]. In Fig. 3(b), the $V_{GS}$ at the kink in the $I_D$ vs. $V_{GS}$ curve is the voltage at which the Fermi energy is closest to the trap energy, and charges a significant amount of traps. For large $V_{GS}$, all traps are charged, and the drain current remains constant for different trap energies. Similarly, Fig. 4(a) shows the impact of doping the channel material for an n-type 2D FET. Large doping shifts the flatband voltage in the negative direction, increasing the current.

### B. Mobility

The electron mobility in 2D materials depends on the vertical and the lateral electric field, as well as on the temperature. The dependence on *vertical* (gate) field comes in through the dependence on carrier density. Higher carrier density can partially screen scattering with ionized impurities and remote polar phonons,[28] and higher carrier density also raises the Fermi level, which changes the effective density of states for scattering. Classical "sixth-power law of thickness" surface roughness scattering present in ultra-thin (<3 nm) bulk semiconductors like Si[29] should not, in principle, affect 2D semiconductors without dangling bonds.[28] However, the vertical field could affect scattering with microscopic roughness of the gate dielectric, including the remote phonons mentioned above. The mobility dependence on *lateral* field mostly comprises high-field effects, i.e. drift velocity saturation. The temperature dependence of mobility comes in through scattering with intrinsic phonons (of the 2D material) and remote dielectric phonons.

Keeping the above considerations in mind, we fit the mobility behavior (at low lateral field) with the following semi-empirical relationship:

$$\mu_{eff} = \frac{\mu_0}{\left(1 + \dfrac{F_V}{F_C}\right)^{\eta} \left(\dfrac{T}{T_0}\right)^{\gamma}}, \tag{7}$$

where $\mu_0$ is the effective mobility at zero field and room temperature ($T_0$), $T$ is the average device temperature, $F_V$ is the vertical electric field, $\gamma$ is a positive constant that depends on dominant phonons, and fitting parameters $\eta$ and $F_C$ depend on the material and the quality of the interface. Figure 4(b) compares this model with experimental data for 1L WSe$_2$ (Ref. 27) and the fit provides $\eta = 6.8$ and $F_C = 305$ V/µm. Previous studies on 1L MoS$_2$ have observed $\eta = 1.45$ and $F_C = 90$ V/µm.[14] The value of $\gamma$ is also obtained by fitting the model to experimental data, yielding $\gamma$ (bulk) = 2.6 (Ref. 30) and $\gamma$ (1L) = 1 to 1.6 for electron mobility in MoS$_2$.[3,4]

At high lateral field, the carrier drift velocity begins to saturate and the effective mobility decreases. We include this effect using a semi-empirical relation

$$\mu = \frac{\mu_{eff}}{\left[1 + \left(\dfrac{\mu_{eff} F}{v_{sat}}\right)^{\xi}\right]^{1/\xi}} \tag{8}$$

where $\xi$ is a fitting parameter with typical value around 2 to 4, $F$ is the lateral electric field and $v_{sat}$ is the saturation velocity. Note it is this $\mu$ which is then used when calculating the current in eq. 6. The temperature dependence of the saturation velocity is incorporated similarly to models for graphene and Si,[31,32] as $v_{sat} = v_0/(1 + N_{OP})$ where the OP (optical phonon) occupation is $N_{OP} = 1/[\exp(\hbar\omega_{OP}/(k_B T)) - 1]$. Here $\hbar\omega_{OP}$ is the OP energy and $v_0$ can be interpreted as the saturation velocity extrapolated to zero Kelvin. For MoS$_2$ the best fit against experimental high-field data (on monolayer MoS$_2$ grown by chemical vapor deposition on SiO$_2$) is obtained with $\hbar\omega_{OP} \approx 30$ to 40 meV and $v_0 \approx 2$ to $3 \times 10^6$ cm/s.[33] However, when modeling *I-V* curves of different devices in the literature, we must treat $v_0$ and $\hbar\omega_{OP}$ as fitting parameters.

### C. Self-Heating Model

Considering that 2D devices can carry high current densities,[3,34,35] unlike their organic counterparts in flexible and transparent electronics,[36] such transistors can generate significant heat. We can model the FET self-heating by including a thermal resistance ($\mathcal{R}_{TH}$) such that the average device temperature rise is $\Delta T = T - T_0 = P\mathcal{R}_{TH}$, where $P = I_D(V_{DS} - 2I_D R_C)$ is the power input of the device without the contacts. As illustrated in the inset of Fig. 5(a), the total thermal resistance has three components: the thermal boundary resistance (TBR) between the channel and the bottom oxide ($\mathcal{R}_B = \mathcal{R}_{TBR}/A$), the spreading resistance of the bottom oxide ($\mathcal{R}_{BOX}$), and the spreading thermal resistance into the substrate ($\mathcal{R}_{Si}$).[37,38] The thermal resistance per unit length is given as

$$g^{-1} = \frac{\mathcal{R}_{TBR}}{W} + \left\{ \frac{\pi k_{BOX}}{\ln\left[6(t_{BOX}/W+1)\right]} + \frac{k_{BOX}W}{t_{BOX}} \right\}^{-1}$$
$$+ \frac{1}{2k_{sub}} \left( \frac{L_G + 2L_U}{W_{eff}} \right)^{1/2}. \tag{9}$$

Here $k_{BOX}$ and $t_{BOX}$ are the thermal conductivity and thickness of the bottom oxide (BOX), respectively, and $k_{sub}$ is the thermal conductivity of the substrate. The TBR per unit area is $\mathcal{R}_{TBR} \approx 10^{-7}$ m$^2$KW$^{-1}$ for monolayer MoS$_2$ on SiO$_2$[39] and the "thermal area" of the device is $A \approx W(L_G + 2L_U)$. Due to heat spreading effects in the SiO$_2$, the effective thermal width at the SiO$_2$/Si interface can be approximated as $W_{eff} \approx W + 2t_{BOX}$.

The thermal expression in eq. 9 strictly only applies to "longer" channel devices, at least three times longer than the lateral thermal healing length ($L_H$) along the channel. For 1L MoS$_2$ on 90 nm SiO$_2$ on Si substrate, $L_H = [k_{2D}t_{2D}(W/g + \mathcal{R}_{TBR})]^{1/2} \approx 100$ nm, if we take $k_{2D} \approx 80$ Wm$^{-1}$K$^{-1}$ as the in-plane thermal conductivity of MoS$_2$ at room temperature. ($k_{2D}$ could become a function of length in shorter devices.[40]) For "longer" devices (with $L_G + 2L_U > 3L_H$), the thermal resistance is given simply by $\mathcal{R}_{th} \approx 1/[g(L_G + 2L_U)]$. For "shorter" channel length devices ($L_G + 2L_U < 3L_H$), heat flow into the metal contacts becomes non-negligible and can be taken into account as described in Ref. 38. We note that heat flow into a top metal gate can, in general, be neglected, partly due to TBR at the two top oxide interfaces, but mainly due to the presence of the larger heat sink (i.e. Si substrate) at the back-side.[41]

To quantify the impact of device self-heating, we simulate a typical 1L MoS$_2$ transistor ($L_G$ = 1 μm) in Fig. 5(b) under four circumstances: without self-heating (solid line, top curve), with self-heating on 90 nm SiO$_2$ / Si substrate (dashed), with self-heating on 300 nm SiO$_2$ / Si substrate (dotted), and finally on a poor thermal substrate where the Si was replaced by a plastic like polyethylene naphthalate (PEN) or acrylic (dash-dotted). We observe reduction in saturation current of approximately 20, 26, and 70%, respectively from the "ideal" scenario without any self-heating. Thus self-heating becomes crucial for devices on substrates with poor thermal conductivity, especially when the FET channel is a high-quality 2D material which has large current-carrying capability.

### D. Capacitance Modeling

We now discuss the key parasitic capacitances that contribute to $C_{GS}$ (gate to source) and $C_{GD}$ (gate to drain). The total parasitic capacitance between the gate and the channel nodes (source and drain) is due to internal fields through the channel ($C_{if}$), outer fringing fields through the surrounding region ($C_{of}$) and normal fringing fields between the gate, and the source or the drain ($C_{nf}$). We display these fields in the schematic shown in inset of Fig. 5(a). The S2DS model does not include the capacitance between the gate and metal plugs at the drain or the source, but such capacitance can easily be included following Ref. 42.

The capacitance $C_{nf}$ is obtained by mapping the perpendicular surfaces of the gate sidewall and the top surface of contact metal to equivalent parallel surfaces using conformal mapping.[43] We modify $C_{nf}$ from Ref. 43 to only include the part of the gate sidewall ($t_G + t_{OX} - t_C$) which is higher than the contact metal as shown in the Fig. 5(a) inset. We assume that the contact length ($L_C$) is larger than the underlap length ($L_U$). $C_{of}$ includes the fringing fields from the horizontal edges of the gate metal to the horizontal edges of the contact metal.[43] In addition, $C_{of}$ includes the parallel capacitance between vertical sidewalls of the gate and the source or drain, approximated with an average distance ($L_U^2 + t_{OX}^2$)$^{1/2}$ between sidewalls. By solving for the specific geometry in Fig. 5(a), we obtain the analytical form of the total fringing capacitance as:

$$C_{nf} = \frac{2W\varepsilon_{sp}}{\pi} \ln\left[\frac{1.3(t_G + t_{ox} - t_C) + \sqrt{L_U^2 + 1.3^2(t_G + t_{ox} - t_C)^2}}{L_U}\right] \quad (10a)$$

$$C_{of} = \frac{0.2W\varepsilon_{sp}}{\pi} \ln\left[\frac{\pi W}{\sqrt{L_U^2 + t_{OX}^2}}\right] \exp\left(-\left|\frac{L_U - t_{OX}}{L_U + t_{OX}}\right|\right) + \frac{W(t_G + t_C)\varepsilon_{sp}}{2\sqrt{L_U^2 + t_{OX}^2}} \quad (10b)$$

where $\varepsilon_{sp}$ is the dielectric constant of the surrounding spacer region, and the other quantities are defined in Fig. 1 and Fig. 5(a).

To obtain $C_{if}$, we separate the contribution of the channel charge between the source and the drain terminals by using the Ward-Dutton charge partition scheme.[44]

$$Q_D = W \int_0^{L_G} \frac{x}{L_G} Q_n(x) dx \quad (11a)$$

$$Q_S = W \int_0^{L_G} \left(1 - \frac{x}{L_G}\right) Q_n(x) dx \quad (11b)$$

Here the charge at the source ($Q_S$) and the drain ($Q_D$) are written in terms of the position dependent channel charge $Q_n(x) = -qn_{2D}(x)$. We note, as described in section II-A, that $n_{2D}(x) = N_{2D} \ln(1 + \alpha)$ where $\alpha = \exp[(qV_C(x) - E_0)/(k_B T)]$, and additional details are discussed in Appendix B. Using nodal charges, we calculate the internal field capacitances between node $m$ and node $j$ as $C_{if} = -\partial Q_m/\partial V_j$ ($m \neq j$) where $m$ and $j$ are the transistor nodes (gate, source or drain).

We also consider the impact of the fringing field from the top-gate on the carriers in the underlap region by including an effective fringing capacitor ($C_{tf}$) from the gate metal sidewall to the underlap region. The analytical expression for $C_{tf}$ is obtained similar to eq. 10 using conformal mapping:[43]

$$C_{tf} = \frac{2W\varepsilon_{sp}}{\pi} \ln\left[\frac{\phi t_G + t_{ox} + \sqrt{\phi^2 t_G^2 + 2\phi t_G t_{ox}}}{t_{ox}}\right] \quad (12a)$$

$$\phi = \exp\left(\frac{L_U - \sqrt{t_G^2 + 2t_{ox}t_G}}{3.7 L_U}\right) \quad (12b)$$

We note that when $t_{OX} \ll t_G \approx L_U$, then $\phi \approx 1$.

### E. Extrinsic Resistance

Generally speaking, the total extrinsic resistance $R_{ext} = R_U + R_C$ of the intrinsic device includes resistance due to the underlap region ($R_U$) and the contact resistance between metal and semiconductor ($R_C$).The underlap $R_U$ can be reduced by adjusting the back-gate voltage (i.e. "electrostatic doping" using a back-gate plane under the entire device), or by chemical doping, the latter being preferred in realistic devices. The impact of both adjustments is shown in Fig. 6(a) and Fig. 6(b) for different underlap lengths. In addition, the underlap resistance can also be reduced by increasing the mobility of the 2D channel material.

The contact resistance ($R_C$) for 2D devices can display non-linear behavior with respect to drain and gate voltages due to the Schottky barrier at the metal-semiconductor interface.[7] For simplicity, here we assume that $R_C$ is optimized in the fabrication and is Ohmic, which is a good approximation at higher lateral field and high $V_{DS}$. Nonetheless, following work on organic TFTs,[45,46] it is possible to model each contact with a pair of anti-parallel Schottky diodes.

In Fig. 6(c) we plot the fringe capacitance $C_F (= C_{of} + C_{nf})$ for different dielectric constants of the spacer region. As expected, the capacitance is largest for small $L_U$ and drops off to a small value for larger underlap lengths. For this particular device geometry (see figure caption), we plot $R_U$ and $C_F$ in Fig. 6(d). To minimize the $R_U C_F$ product, the optimal $L_U \approx 25$ nm in this case. If the underlap resistance is reduced (e.g. by higher doping or higher mobility in this region), the optimized underlap length will increase. Another solution to reduce the optimized $L_U$ is to decrease the dielectric constant of surrounding material, $\varepsilon_{sp}$.

### III. MODEL CALIBRATION

We calibrate the model to monolayer 2D semiconductors including $MoS_2$ (n-type) and $WSe_2$ (p-type) using existing experimental data. We note that this calibration is not universal, i.e. it can be improved or adjusted as more data become available from improved nanofabrication of such devices. The model is also not restricted to monolayer devices, although thicker devices could require additional treatment of depletion capacitance[26] and band structure. Here, we use monolayer $MoS_2$ devices of various lengths ranging from $L_G \approx 80$ nm to a few μm in order to calibrate the model.[3,11] The device contact resistance (e.g. ~1 to 6 k$\Omega$·µm depending on doping and back-gate voltage) and the mobility (20-40 cm$^2$V$^{-1}$s$^{-1}$) are independently measured through experimental transfer line method (TLM), reassuring our fit.[3]

The model provides a good fit for long-channel back-gated [Fig. 7(a)] as well as top-gated [Fig. 7(b)] monolayer MoS$_2$ devices. We also fit the model to long-channel 1L WSe$_2$ (Ref. 27) and extract effective $\mu_0$ = 245 cm$^2$V$^{-1}$s$^{-1}$, device doping ($N_{\text{Dop}} \sim 10^{13}$ cm$^{-2}$), and trap density ($D_{\text{it}} \sim 10^{12}$ cm$^{-2}$). The model comparison to $I_\text{D}$-$V_\text{DS}$ and $I_\text{D}$-$V_\text{GS}$ experimental data for 1L WSe$_2$ p-type FET are shown in Fig. 7(c) and Fig. 7(d).

We note that the S2DS model can reproduce all experimental (*I*-*V*) data sets we have examined (from our lab and the published literature) with approximately ten fitting parameters. These include $\mu_0$, $\beta$, $\xi$, $V_{\text{GS0}}$, etc., as described above. Other material parameters like carrier effective masses,[22] thermal conductivities,[40] and band structure,[19–21] are generally imported from first principles simulations or published experimental data and are not treated as fitting parameters. Nevertheless, these are available in the model and can be adjusted as future data become available (e.g. influence of strain on effective masses, etc.).

## IV. VERILOG-A AND CIRCUIT SIMULATIONS

The S2DS compact model is implemented in Verilog-A (Ref. 16) to perform circuit simulations in SPICE, and the code is freely available online.[17] Standard modeling guidelines were followed to capture the device behavior in Verilog-A.[47] Device self-heating is currently modeled in DC mode only, but time-dependent heating could also be modeled by including the device thermal capacitance, which will be dominated by the materials surrounding the 2D channel rather than the 2D channel itself.[41] We utilize the multi-physics support in Verilog-A to include the lumped self-heating model using a thermal node. We use additional internal nodes to calculate the quantum potentials near the source and the drain ($V_{\text{Cs}}$ and $V_{\text{Cd}}$).

Before concluding, we wish to illustrate the use of the S2DS model in simple circuit simulations, such as the inverters and ring oscillators in Fig. 8. (Additional system-level simulations, including the role of variability, will be the scope of future work.) For these circuit demonstrations we calibrate n-type devices to 1L MoS$_2$ and p-type devices to 1L WSe$_2$. For MoS$_2$ we use $\mu_0$ = 81 cm$^2$V$^{-1}$s$^{-1}$ (Ref. 48) and for WSe$_2$ we use $\mu_0$ = 245 cm$^2$V$^{-1}$s$^{-1}$ (Ref. 27). For this test case, we consider channel lengths $L_G = 100$ nm, effective oxide thickness (EOT) = 2 nm, and contact resistance $R_C = 200 \ \Omega \cdot \mu m$. Over the range of gate voltages applied (supply voltage $V_{DD} = 1$ V), the mobility ratio of $WSe_2$ to $MoS_2$ is 3:1. Thus, we used a width ratio $W_p/W_n = 1:3$ to balance the inverter, where $W_p = 1 \ \mu m$ and $W_n = 3 \ \mu m$ are widths of FETs from $WSe_2$ and $MoS_2$, respectively. The saturation drive currents of the optimized devices are approximately ~350 µA.

From our data-calibrated model, we predict a DC inverter gain of ~130, as seen in Fig. 8(a). We also simulate a 3-stage ring oscillator [inset of Fig. 8(b)] in SPICE using the above inverter designs. We load the individual inverters with a 30 fF load ($C_L$) to emulate the interconnect capacitance. The period of the oscillator is close to 0.5 ns for $V_{DD} = 1$ V as seen in Fig. 8(b). These results are not fundamental, but reliant on the particular test case chosen here, to illustrate the practicality of the S2DS model. Further improvements in mobility, drive currents, tailoring of flat-band voltages ($V_{GS0}$ and $V_{BS0}$) using different gate metals, and optimization of device dimensions (e.g. $L_U$) will further help to improve the delays and DC gain.

## V. CONCLUSION

In summary, we described S2DS, a physics-based, predictive 2D transistor model capable of simulating devices and circuits from 2D semiconductors like $MoS_2$ and $WSe_2$. S2DS is exclusively geared toward 2D materials, which cannot be readily treated with existing models for ultra-thin body (UTB) silicon on insulator (SOI).[49–51] S2DS includes the quantum capacitance of 2D monolayers, anisotropic 2D material properties (e.g. thermal anisotropy), as well as the 2D band structure and mobility calibrated with existing experimental data. S2DS also includes a thermal model, which is important for self-consistent high-field simulations, illustrating that 2D-FET saturation currents are thermally limited, especially on thermally insulating (e.g. flexible, plastic) substrates. The model is presented with sufficient generality that it can be easily updated as additional data become available for 2D materials and devices.

We also demonstrate the capability of S2DS to extract and quantify unknown variables such as doping density and trap density from electrical data. The model can be used to optimize device dimensions like the gate-source underlap length, as well. Importantly, S2DS is implemented in Verilog-A, and the code is freely available on the nanoHUB.org.[17] All the equations are analytical and can be comfortably used in the SPICE environment for multi-transistor circuit simulations. Such a compact model is a major step towards developing systems from 2D materials, and it could also be used to benchmark and select various 2D materials with future systems in mind.

## ACKNOWLEDGMENT

The authors would like to thank Kirby Smithe, Chris English, Shaloo Rakheja and Chi-Shuen Lee for their experimental data and useful discussion. This work was supported in part by the NCN-NEEDS program, which is funded by the National Science Foundation, contract 1227020-EEC, and by the Semiconductor Research Corporation (SRC). This study was also partly supported by the NSF EFRI 2-DARE grant 1542883, and by Systems on Nanoscale Information fabriCs (SONIC), one of the six SRC STARnet Centers, sponsored by MARCO and DARPA.

## APPENDIX

### A.  Derivation of I-V Characteristics

We solve for the dependence of channel charge ($Q_{ch}$, eq. 3), bias potentials, and oxide capacitances as follows:[13,24]

$$V_C(x) = \frac{Q_{ch}(V_C)}{C_t + C_b} + \left[V_{GS} - V_{GS0} - V_n(x)\right]\frac{C_t}{C_t + C_b} + \left[V_{BS} - V_{BS0} - V_n(x)\right]\frac{C_b}{C_t + C_b} \ . \quad (A1)$$

The above equation solved iteratively with eq. 3 provides the quantum potential $V_C$ at the drain [calculated at $V_n(x) = V_{DS}$] and at the source [calculated at $V_n(x) = 0$]. $C_t$ and $C_b$ are the capacitances of the top and bottom oxide, respectively. $V_{GS}$ and $V_{BS}$ are the voltages at the top-gate and the back-gate with respect to the source terminal. $V_{GS0}$ and $V_{BS0}$ are the flat band voltages of the top- and back-gate, respectively, for an undoped channel, and are treated as fitting parameters. $V_n(x)$ and $V_C(x)$ are the channel potential and the quantum potential respectively. Differentiating both sides of eq. A1 with respect to $V_C(x)$, we obtain

$$\frac{dV_C(x)}{dV_n(x)} = -\left(1 + \frac{C_q + C_{it}}{C_t + C_b}\right), \quad (A2)$$

where $C_q$ and $C_{it}$ are provided in eq. 5. We assume drift transport in the channel,

$$I_D = Q_n(x)v(x)W = Q_n\big[V_n(x)\big]v[V_n(x)]W \ . \tag{A3}$$

By substituting $F = -dV_n(x)/dx$ and integrating the equation over $x$ from $[0, L_G]$ we obtain

$$I_D = \mu W \int_0^{L_G} F Q_n\big[V_n(x)\big] dx \tag{A4}$$

where we have assumed the mobilty to be constant along the channel. After changing the variable from $x$ to channel potential $V_n$, the above equation is re-written as,

$$I_D = \mu \frac{W}{L_G} \int_{V_{Cs}}^{V_{Cd}} Q_n\big(V_n\big) \frac{dV_n}{dV_C} dV_C \tag{A5}$$

Note that $Q_n(V_n)$ $[= -qn_{2D}(V_n)]$ is the charge due to the free carriers (electrons in this case). $V_{Cd}$ and $V_{Cs}$ are the quantum potentials at the drain and source respectively. The final equation is derived by substituting capacitance expressions from eq. A2 and eq. 5. After simplifying, we obtain the following analytical form of the drain current.

$$I_D = \frac{\mu W}{L_G}\left[ N_{2D}k_BT\alpha + q^2\ln^2(1+\alpha) - \frac{N_{2D}q^2D_{it}\beta}{(C_t+C_b)(\beta-1)}\left( \frac{(1+\alpha)\ln(1+\alpha)}{\alpha+\beta} - \ln(\alpha+\beta) \right) \right]_{V_{Cd}}^{V_{Cs}} \tag{A6}$$

### B.  Calculation of the channel charge partition

We use the Ward-Dutton partition[44] to simplify the expressions in eq. 11:

$$Q_D \cong Q_{n0}\frac{4+8\theta+12\theta^2+6\theta^3}{15(1+\theta)^2} \tag{B1}$$

$$Q_S \cong Q_{n0}\frac{6+12\theta+8\theta^2+4\theta^3}{15(1+\theta)^2} \tag{B2}$$

$$Q_{n0} \cong -qWL\frac{n_{2D}(x=0)+n_{2D}(x=L_G)}{2} \tag{B3}$$

$$Q_G = -\big(Q_S + Q_D\big) \tag{B4}$$

The above expressions (B1 and B2), although derived for bulk MOSFETs,[44] have been previously used to calculate charge at the drain and the source nodes in carbon nanotube FETs[52] as well as graphene FETs.[53] The symbol $\theta$ is an empirical function of $V_{DS}$, which is used to provide a continuum between the different regions of transistor operation. From the linear regime $\theta = 1$ to the saturation regime $\theta = 0$.[53] $\theta = 1 - F_q$ such that $F_q = (V_{DS}/V_{DSsat})/[(V_{DS}/V_{DSsat})^a + 1]^{1/a}$, with a typical value $a = 3$ to 4, to make $\theta$ continuous. $V_{DSsat}$ is the saturation voltage at which the channel charge near the drain becomes zero or $n_{2D}$(at $x = L_G$) = 0. It is approximated as $V_{DSsat} = Q_{n0}/[WL_GC_t(1+\omega)]$. $\omega$ is a fit parameters ranging from 0 to 1.[44]

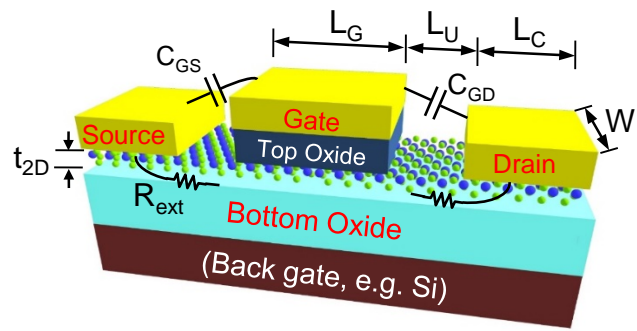

**Figure 1**: Schematic of a representative 2D semiconductor FET, including its parasitic elements. Here, the channel is a monolayer semiconductor such as $MoS_2$. The channel thickness and the width are represented by $t_{2D}$ and $W$ respectively. $L_G$ is the gate length, $L_U$ is the underlap length, and $L_C$ is the length of the source and drain contacts. $C_{GS}$ and $C_{GD}$ are the capacitances from gate to source and to drain, respectively. $R_{ext}$ is the total external resistance that includes the contact resistance ($R_C$) and the resistance of the underlap region ($R_U$). The substrate can be doped Si, functioning as a back-gate, or it could be an electrically insulating polymeric flexible substrate.

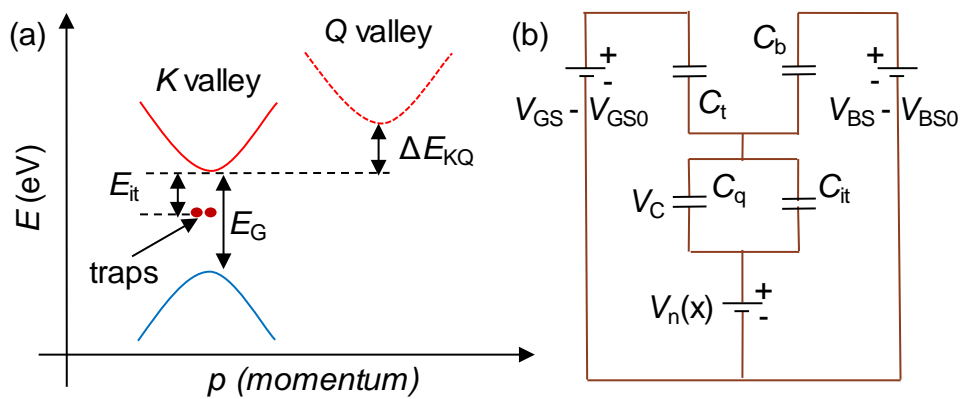

**Figure 2**: (a) Schematic of band structure of a 2D material with the K and the Q conduction band valleys. The energy separation between two valleys, $\Delta E_{KQ}$, is of the order of few $k_B T$. (b) Schematic used to calculate the channel charge. $V_{GS}$ and $V_{BS}$ are the voltages of the top- and back-gate, respectively. $V_{GS0}$ and $V_{BS0}$ are respective flat band voltages. $C_t$ and $C_b$ are the top and bottom oxide capacitances, respectively. $C_q$ is the quantum capacitance of the 2D monolayer channel and $C_{it}$ is the capacitance due to traps at the oxide-channel interface(s).

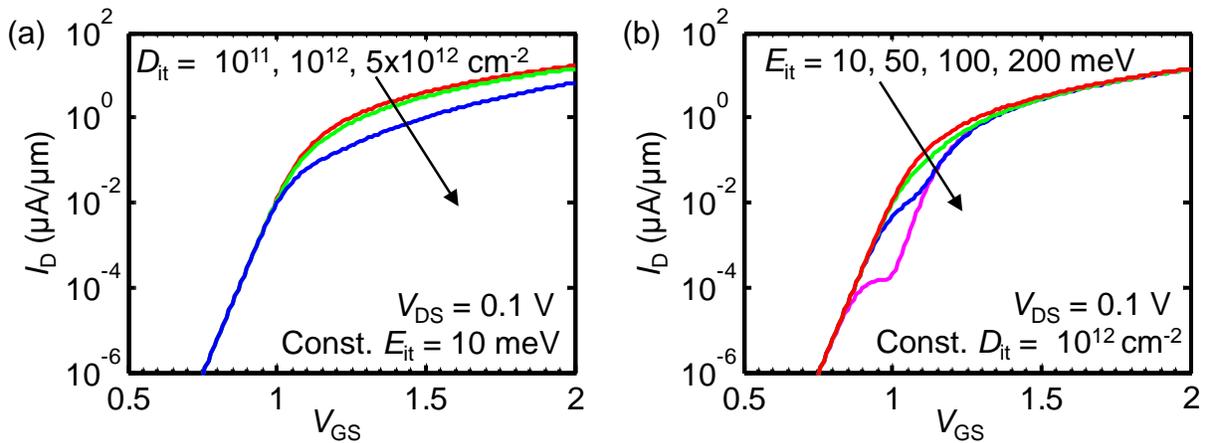

**Figure 3**: Simulated drain current vs. top-gate voltage ($I_D - V_{GS}$) curves for a 1L top-gate MoS$_2$ n-FET ($V_{BS} = 0$) with $L_G = 1.0$ μm, effective oxide thickness (EOT) = 2 nm, $R_{ext} = 1$ kΩ·μm and $\mu_0 = 80$ cm$^2$V$^{-1}$s$^{-1}$ with (a) varying trap density ($D_{it}$); here $E_{it} = 10$ meV and (b) varying trap depths ($E_{it}$); here $D_{it} = 10^{12}$ cm$^{-2}$. For larger gate bias all traps are charged, and we see that the current is the same irrespective of $E_{it}$.

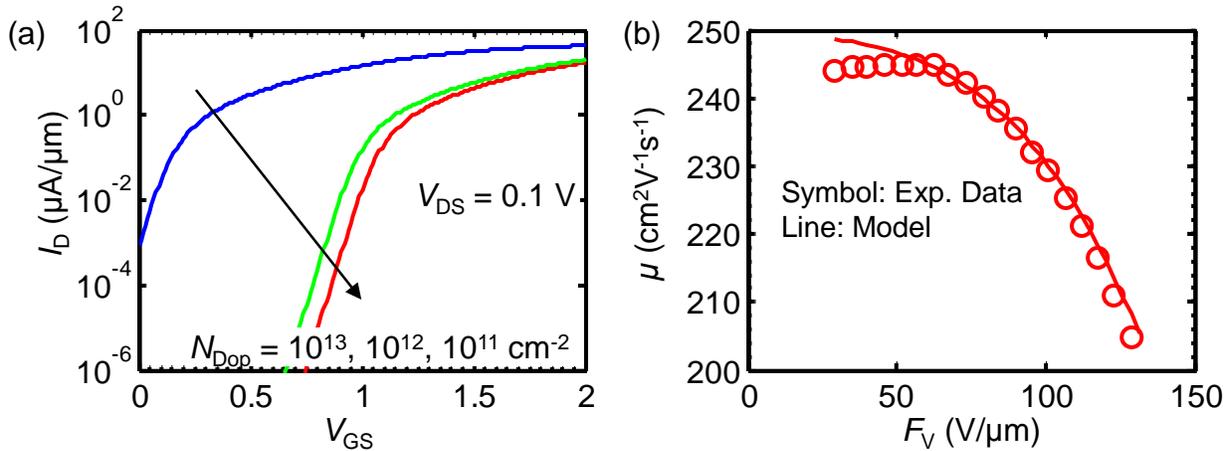

**Figure 4**: (a) Simulated $I_D - V_{GS}$ of a 1L single-gate MoS$_2$ n-FET with the same characteristics as in Fig. 3, but with varying doping density and $D_{it} = 0$. Channel doping changes the flatband (and consequently the threshold) voltage of the device. (b) Calibration of the mobility model with experimental data for 1L WSe$_2$, showing dependence on vertical electric field ($F_V$).

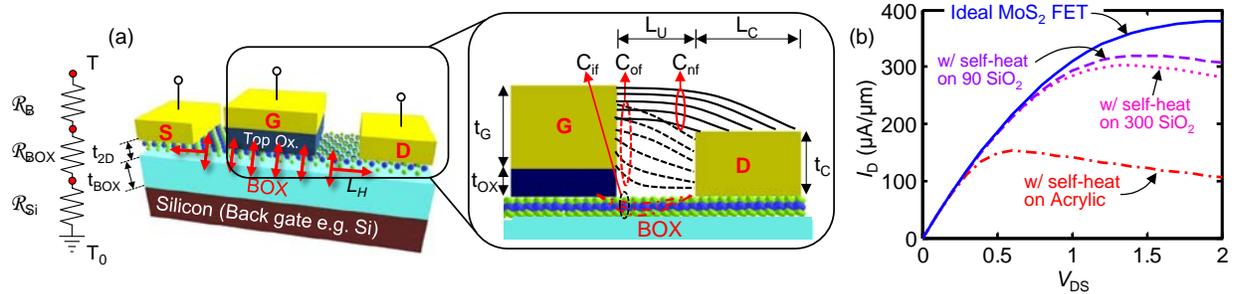

**Figure 5**: (a) Schematic representation of heat flow (red lines) in a 2D FET and associated lumped thermal model. $T$ is the average device temperature and the substrate bottom is at the ambient $T_0$. The inset displays a schematic diagram showing all fringe capacitances included in the model. (b) Simulated drain current vs. drain voltage ($I_D - V_{DS}$) for an "ideal" MoS₂ transistor without self-heating (solid lines, top curve), with self-heating on 90 nm SiO₂ / Si substrate (dashed), with self-heating on 300 nm SiO₂ / Si substrate (dotted), and on a poor thermal substrate where the Si was replaced by PEN or acrylic (dash-dotted, $k_{acr} \approx 0.2$ Wm⁻¹K⁻¹). The simulation assumes $W = 1$ μm, $L_G = 1$ μm, $L_U = 100$ nm, $\mu_0 = 80$ cm²V⁻¹s⁻¹, $V_{GS} = 2$ V, top EOT = 2 nm, $R_{ext} = 1$ kΩ·μm and $V_{BS} = 0$ V.

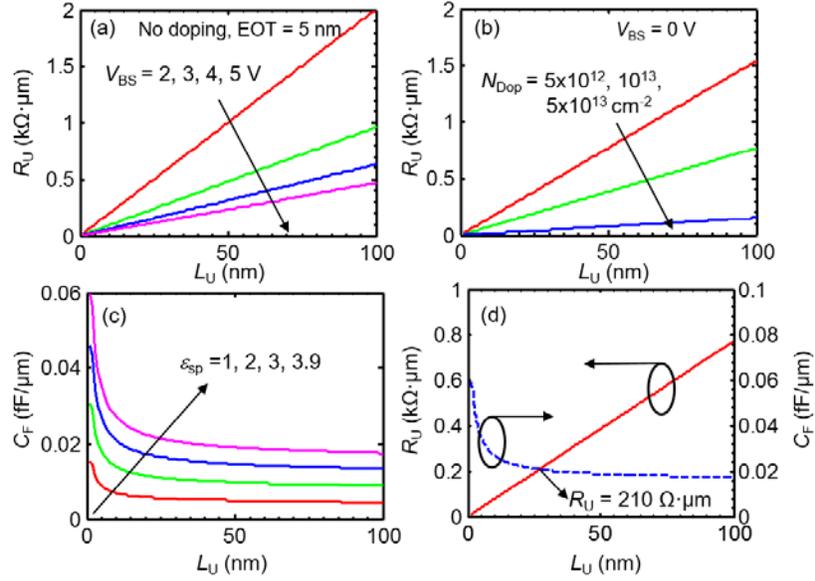

**Figure 6**: Using the S2DS model to optimize the underlap resistance. The geometry in Fig. 1 is assumed, with a global $p^+$ silicon back-gate ($V_{BS0} = 0.22$ V), EOT = 5 nm for the back oxide, and $\mu_0 = 80$ cm²V⁻¹s⁻¹. All values are calculated at $V_{DS} = 0$ V. (a) The underlap resistance is plotted for $V_{BS} = 2$ to 5 V. (b) Effect of doping in reducing the resistance of the underlap regions. (c) The fringe capacitance ($C_F = C_{of} + C_{nf}$) as function of underlap length for different dielectric constants. Here $t_C = t_G = 40$ nm and $t_{OX} = 2$ nm. (d) Trade-off between underlap resistance and fringe capacitance for a nominal case with $V_{BS} = 0$ V, $N_{Dop} = 10^{13}$ cm⁻² and $\varepsilon_{sp} = 3.9$.

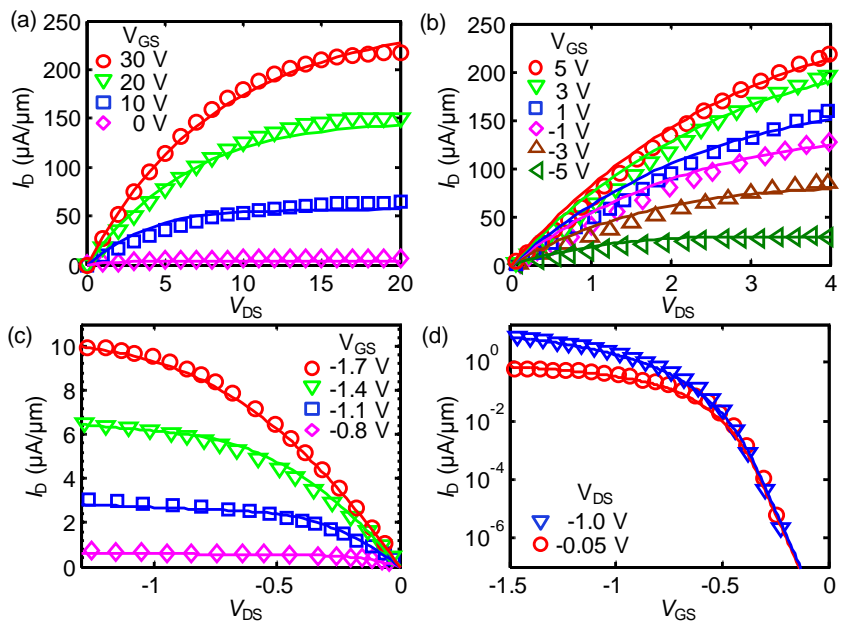

**Figure 7**: S2DS model calibration for monolayer 2D devices. Simulations are displayed with lines, experimental data are symbols. (a) Simulated $I_D$ – $V_{DS}$ compared to experimental data for $L_G$ = 3.2 µm back-gated, CVD-grown MoS₂.[3] (b) Simulated $I_D$ – $V_{DS}$ vs. experimental data for $L_G$ = 250 nm top-gated MoS₂ device.[11] We fit monolayer WSe₂ data[27] ($L_G$ = 9.4 µm) with (c) $I_D$ – $V_{DS}$ for different top-gate voltages and (d) $I_D$ – $V_{GS}$ for different drain voltages.

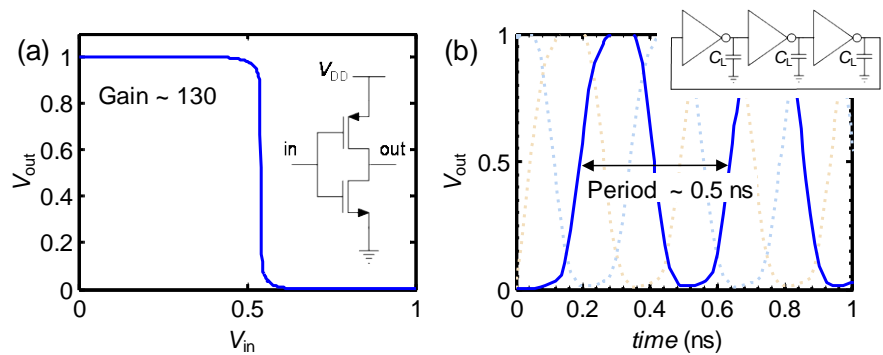

**Figure 8:** (a) Inverter sweep for $V_{DD}$ = 1 V. Inset shows schematic of an inverter with complementary p-FET (WSe$_2$) and n-FET (MoS$_2$) with $L_G$ = 100 nm. (b) Output of a 3-stage ring oscillator with a period of ~ 0.5 ns, at $V_{DD}$ = 1 V. The inset shows a schematic of the 3-stage ring oscillator. The load capacitance $C_L$ = 30 fF.